\documentclass{elsart}
\usepackage{amsmath,amssymb,epsfig}

\begin{document}

\begin{frontmatter}

\title{The structure of nuclear systems derived from  \\
low momentum nucleon-nucleon potentials}
\author{J. Kuckei, F. Montani, H. M\"uther and  A. Sedrakian}
\address{ Institut f\"ur Theoretische Physik, Universit\"at T\"ubingen,
D-72076 T\"ubingen, Germany}
\date{\today}


\begin{abstract}
Various nuclear structure observables are evaluated employing low-momentum
nucleon-nucleon (NN)  potentials $V_{\rm low-k}$  derived from the CD-Bonn and
Nijmegen NN interactions  $V_{NN}$.  By construction, the high momentum modes
of the original $V_{NN}$  are integrated out in $V_{\rm low-k}$, with the
requirement that  the deuteron binding energy and low energy phase shifts of
$V_{NN}$ are exactly reproduced.  Using this interaction, we evaluate the bulk 
properties (binding energy and saturation density) of nuclear matter  and
finite nuclei, in particular their dependence on the cut-off parameter.  We
also study the pairing gap and the residual interaction in nuclear matter in
terms of the Landau parametrization. At low and medium densities, the HF and
BHF binding energies for nuclear matter  calculated with the $V_{\rm low-k}$'s
derived from the CD-Bonn and  Nijmegen potentials are nearly identical. The
pairing gaps and Landau  parameters derived from  $V_{\rm low-k}$ are
remarkably close to those  given by the  full-space $V_{NN}$. The $V_{\rm
low-k}$ interactions,  however, fail to reproduce the saturation property of
nuclear matter at higher densities if the cut-off for the high momentum modes
is assumed density independent.


\vspace{0.5cm}

\noindent{\it PACS:}
21.30.Fe;          
13.75.Cs;	   
27.80.+j;          
21.60.-n;          
\\	
\noindent{\it Keywords:}  Forces in hadronic systems and effective 
interactions; Nucleon-Nucleon Interactions; Properties of nuclei;
Nuclear-structure: models and methods
\end{abstract}
\end{frontmatter}

\section{Introduction}
In the past decade, there has been much progress in treating
low-energy nuclear systems using the effective field theory 
(EFT) and  renormalization group (RG) approaches, 
 \cite{weinberg,kaplan98,lepage,shankar,EFT,kolck,epel98,haxton} and
references therein.
The fundamental theory of strong interaction, QCD, is non-perturbative
in the energy regime characteristic for low energy nuclear physics.
This poses a major difficulty for a proper derivation of the nuclear
force from the first principles. However, 
 this difficulty can be overcome by the EFT-RG approach where
one "integrates out" the fast fields beyond the chiral symmetry breaking scale
$\Lambda _{\chi} \approx 1$ GeV. Well below this scale, the appropriate degrees
of freedom for nuclear systems are pions and nucleons.
In this way, one obtains for these degrees of freedom a chiral
symmetric EFT, which is used to construct either perturbative
or non-perturbative controlled approximations for low energy nuclear systems.

As is well known, there are several realistic models
for the NN potential $V_{NN}$, such as the CD-Bonn  \cite{cdbonn},
Nijmegen  \cite{nijm1} and the Argonne  \cite{argv18} potentials.
These models all have the same long range tail, namely the one-pion
exchange potential. But the short range
parts significantly differ from each other, despite the 
fact they all fit essentially exactly the
empirical deuteron binding energy and low energy phase shifts up to $E_{lab}
\approx $350 MeV. Clearly the short range (high momentum) parts
of these potentials are model dependent and are different e.g. for 
the local and non-local interaction models.
Employing the EFT-RG ideas, Refs.  \cite{bogner01,bogner02,bogschwenk}  
"integrated out" the high momentum modes of these potentials and 
thereby obtained a low momentum NN potential $V_{\rm low-k}$.
The high momentum modes
of the original $V_{NN}$ are eliminated in $V_{\rm low-k}$
with the requirement that the deuteron binding energy and low 
energy phase shifts of $V_{NN}$ are exactly reproduced.
This is in line with the  central theme of the EFT-RG approach
that physics in the infrared region is insensitive to the details of
the short distance dynamics. 

The purpose of this work is to calculate various nuclear structure
observables for infinite nuclear matter
and finite nuclei starting from the $V_{\rm low-k}$ interaction 
derived for the CD-Bonn and Nijmegen 
potentials \cite{bogner01,bogner02,bogschwenk}.  
In particular, we would like to explore to which extent the results derived
from $V_{\rm low-k}$ differ from the predictions obtained for the underlying
interaction model $V_{NN}$ and the sensitivity of the
results on the cut-off parameter $\Lambda$. For that purpose we evaluate the
ground-state energy of nuclear matter and finite nuclei
taking $^{16}$O as a typical example. Apart from the bulk properties of nuclear
systems we also investigate the predictions for the residual interaction
characterized by the pairing gap $\Delta$ of neutron-neutron or proton-proton
pairing as well as the Landau parametrization of the particle-hole interaction
for low-energy excitations in nuclear matter.

The paper is organized as follows. In section 2 we shortly review the 
construction of the low-momentum $V_{\rm low-k}$ . The methods for the evaluation of
ground-state properties of nuclear matter and finite nuclei as well as the 
corresponding results are presented in sections 3 and 4, respectively. Section 5
contains a discussion of the gap equation and the Landau parameters
characterizing the residual interaction. The conclusions of this study are
summarized in section 6.

\section{Determination of low momentum nucleon-nucleon potentials}

We start with a brief review of the method of carrying out the
high-momentum decimation \cite{bogner01,bogner02,bogschwenk}. 
In accordance with the general definition of a
renormalization group transformation, the decimation must be such that low
energy observables calculated from the full $V_{NN}$ must be reproduced by
$V_{\rm low-k}$, but with all loop integrals cut off at some
$\Lambda$.  To meet this requirement, a $T$-matrix equivalence approach
is used \cite{bogner01,bogner02,bogschwenk}. 
One starts from the half-on-shell $T$-matrix
\begin{equation}
  T(k',k,k^2)
= V_{NN}(k',k)
 + \int _0 ^{\infty} q^2 dq  V_{NN}(k',q)
 \frac{m}{k^2-q^2 +i0^+ } T(q,k,k^2 ) \label{eq:lip1},
\end{equation}
and then defines an effective low-momentum $T$-matrix by
\begin{equation}
  T_{\rm low-k }(p',p,p^2)
= V_{\rm low-k }(p',p)
 + \int _0 ^{\Lambda} q^2 dq  V_{\rm low-k }(p',q)
 \frac{m}{p^2-q^2 +i0^+ } T_{\rm low-k} (q,p,p^2)\label{eq:lip2},
\end{equation}
where $\Lambda$ denotes the momentum space cut-off.
The above $T$-matrices satisfy the condition
\begin{equation}
 T(p',p,p^2 ) = T_{\rm low-k }(p',p, p^2 )\, ;\qquad\mbox{for}\quad ( p',p) 
\leq \Lambda.\label{eq:defv}
\end{equation}
$V_{\rm low-k}$ has been obtained by solving the above equations using the
procedures of Ref. \cite{bogner01,bogner02}. In particular the effects of the
folded diagrams are taken into account to derive an effective interaction 
$V_{\rm low-k}$ which is energy-independent.  

We have performed numerical checks to make sure that the resulting
$V_{\rm low-k}$ does preserve the deuteron
binding energy and low energy phase shifts of the original NN potential.
This has been repeated assuming various cut off parameters $\Lambda$ using the
CD Bonn \cite{cdbonn} and the Nijmegen \cite{nijm1} interaction models for the
bare interaction $V_{NN}$. For both of these interactions we have ignored the
charge dependence and used only the proton-neutron part for all interactions
channels.

\section{Energy of Nuclear Matter}
It is well known that the strong short-range and tensor components of
realistic NN interactions induce strong NN correlations in the wave function
of nuclear many-body system. Therefore rather sophisticated approximation
schemes for the solution of the many-body problem are required to obtain
reasonable results for nuclear structure observables. Such approximation schemes
must account at least for the two-nucleon correlations. Typical
examples are variational approaches using correlated basis
functions \cite{akmal}, the coupled cluster approach \cite{kuem}, the
self-consistent evaluation of Green's functions \cite{whd} and the hole-line
expansion within the Brueckner-Hartree-Fock (BHF) approximation \cite{review}.
 
The effective interaction $V_{\rm low-k}$ has been constructed to account for  NN
correlations with momenta larger than the cut-off momentum $\Lambda$ as they
occur for two interacting nucleons in the vacuum. The question which we now
want to address is whether such a treatment of correlations is sufficient to
allow for simple mean field or Hartree-Fock calculations of the nuclear
many-body system. As a first example we consider 
the equation of state of infinite nuclear matter. 

\begin{figure}
\begin{center}
\epsfig{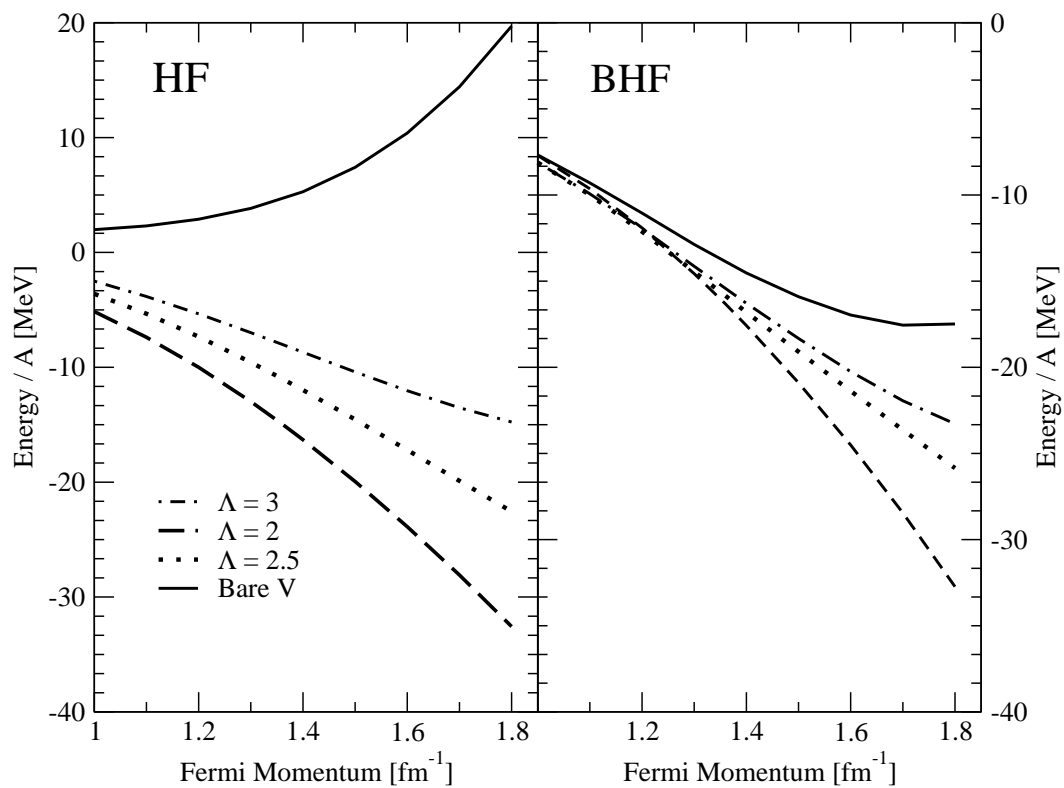}
\end{center}
\caption{Binding energy of nuclear matter as a function of the Fermi momentum
for the HF (left panel) and BHF approximations (right panel). 
The bare interaction (CD-Bonn \protect \cite{cdbonn}) and corresponding 
$V_{\rm low-k}$ interactions with various cut-off parameters $\Lambda$
were used.
\label{fig1}}
\end{figure}

The left panel of Fig.~\ref{fig1} shows the results of Hartree-Fock (HF)
calculations for the energy per nucleon of symmetric nuclear matter as a
function of density, which is characterized in this figure by the corresponding
Fermi momentum $k_F$. For the bare CD-Bonn interaction \cite{cdbonn} the HF
calculations yield positive energies ranging between 3 MeV per nucleon at small
densities and 20 MeV per nucleon for $k_F$ around 1.8 fm$^{-1}$ which
corresponds to a density of around 2.3 times the empirical saturation density.
This demonstrates the well known fact that even modern non-local models for a
realistic NN interaction, which tend to be ``soft'', yield unbound nuclei if the
effects of correlations are ignored \cite{polls}.

This lack of binding energy can be cured if the effects of NN correlations are
taken into account, e.g.~by means of the BHF approximation. In the BHF
approximation one replaces the bare interaction by the $G$-matrix
which obeys the Bethe-Goldstone equation. For a pair of nucleons
with vanishing center of mass momentum it takes the form 
\begin{equation}
G(k',k,2\varepsilon_k) = V_{NN}(k',k)
 + \int _0 ^{\infty} q^2 dq  V_{NN}(k',q)
 \frac{\hat Q}{2\varepsilon_k-2\varepsilon_q +i0^+ } G(q,k,2\varepsilon_k
)\label{eq:betheg} .
\end{equation}
Comparing this equation with the Lippmann-Schwinger equation 
(\ref{eq:lip1}), one can
see that the $G$-matrix is closely connected to the $T$-matrix except for the
fact that the Pauli operator $\hat Q$ in (\ref{eq:betheg}) prevents scattering
into states which are occupied by other nucleons  and that the propagator for the
two nucleons, which for the vacuum is defined in terms of the kinetic energies
(see eq.~(\ref{eq:lip1})) is replaced in (\ref{eq:betheg}) by the propagator in
the medium defined in terms of the single-particle energies $\varepsilon_k$.
Both of these modifications tend to reduce the effects of correlations and are
referred to as Pauli  and dispersive quenching, respectively \cite{rupr}.

It should be mentioned that we employ a very simple version of
the BHF approach, in which the single-particle energies for states below
the Fermi energy are parametrized in terms of an effective mass and a constant
energy shift
\begin{equation}
\varepsilon_k = \frac{k^2}{2m^*} + U\,, \label{eq:mstar}
\end{equation}
while for states above the Fermi energy the single-particle
energies are replaced by the kinetic energy (the so-called conventional
choice).  The Pauli operator is treated in the angle-averaged approximation. A
more  sophisticated treatment of the two-nucleon propagator in the
Bethe-Goldstone equation leads to larger binding energies \cite{khaga1}. Note,
however, that it is not the aim of this study to optimize the BHF
approximation, but to study the possible use of the $V_{\rm low-k}$
interaction in restricted model-space calculations.  

Results of such BHF calculations are displayed in the right panel of
Fig.~\ref{fig1}. The conventional BHF calculation for the bare CD Bonn
interaction yields an energy of  -17 MeV per nucleon and a saturation density 
around twice the empirical saturation density. 
The left panel of Fig.~\ref{fig1} also shows the HF energies calculated for
$V_{\rm low-k}$ assuming three different values for the cut-off parameter
$\Lambda$. The resulting energies are attractive. This demonstrates that the
correlations from two-particle states with momenta $q$ larger than the cut-off
$\Lambda$, which are accounted for in $V_{\rm low-k}$, are sufficient to provide
this attraction. From this consideration it is evident that the energies are
more attractive the smaller the cut-off $\Lambda$, a feature seen 
in Fig.~\ref{fig1}.

Next, in performing BHF calculations for $V_{\rm low-k}$, we must take
into account that $V_{\rm low-k}$ is designed for a model space with relative
momenta smaller than $\Lambda$, i.e. the matrix elements of $V_{\rm low-k}$ must be
supplemented by the value zero for momenta $q$ larger than the cut-off. In other
words, the integral in the Bethe-Goldstone eq.~(\ref{eq:betheg}) is restricted
to momenta $q$ below the cut-off in order to avoid double counting of
the contributions from momenta higher than the cut-off.

The energies resulting from such BHF calculations (see left panel of
Fig.~\ref{fig1}) are below the corresponding HF energies as additional
correlations from NN states above the Fermi momentum but below the cut-off are
taken into account. One also observes that the BHF calculations using
$V_{\rm low-k}$ lead to more attractive energies than the BHF calculation using the
bare interaction. This is due to the fact, that the dispersive quenching
effects in the $G$-matrix discussed above are taken into account only for
intermediate two-particle states with momenta below the cut-off $\Lambda$,
while the BHF calculation using the bare interaction considers these effects
for all states.   
Since the dispersive quenching is a very important ingredient of BHF
calculations, that provides the saturation effect, the BHF calculations employing
$V_{\rm low-k}$ do not exhibit a minimum in the energy versus density curve. One may
conclude that $V_{\rm low-k}$ yields a good approximation for BHF calculations at
small densities. In this regime even a Hartree-Fock calculation leads to
reasonable results, if $\Lambda$ is small enough. At higher densities, however,
$V_{\rm low-k}$ leads to too much binding as it does not account for the dispersive
quenching of high-momentum modes.

\begin{figure}
\begin{center}
\epsfig{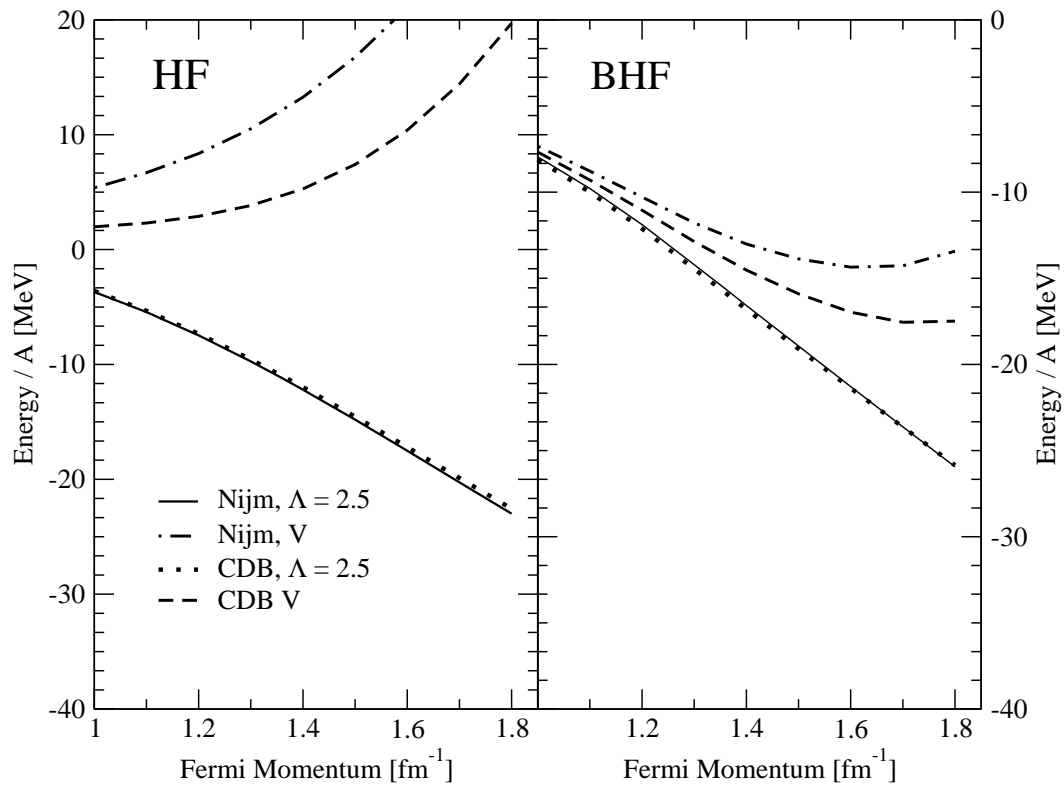}
\end{center}
\caption{Binding energy of nuclear matter as a function of the Fermi
momentum for the HF (left panel) and BHF approximations (right panel).
The interactions are modeled by the CD-Bonn and Nijmegen potentials
and corresponding $V_{\rm low-k}$ interactions with 
a cut-off parameter $\Lambda = 2.5$ fm$^{-1}$.
\label{fig2}}
\end{figure}

In Fig.~\ref{fig2} we compare results derived from two different NN interaction
models: the CD-Bonn (used above)
and the Nijmegen \cite{nijm1} potentials. One can see from
this figure that the HF calculations using the Nijmegen interaction yield
energies which are more repulsive than those derived from the CD-Bonn
interaction. The non-local CD-Bonn is ``softer'' than the local Nijmegen
interaction. As it has been discussed e.g.~in Ref.~ \cite{polls} the
softer interaction model yields more attractive energies also within the BHF
approximation (see right panel of Fig.~\ref{fig2}). Although the results of HF
as well as BHF calculations are rather sensitive to the choice of the
interaction, the corresponding results from $V_{\rm low-k}$ are rather insensitive
to the interaction model on which they are based. This is true for
both the HF and the BHF approximations. 
This can be understood from the fact that the
$T$-matrix is essentially independent on the interaction used, as all realistic
interactions are fitted to the empirical phase shifts. Since the definition of
$V_{\rm low-k}$ is related to the $T$-matrix (see eq.~(\ref{eq:defv})), the
resulting effective interaction is rather insensitive to the original
interaction it is based on.

\section{Finite Nuclei}

Since the use of $V_{\rm low-k}$ yields rather reasonable results for the energy of
nuclear matter at small densities it is of interest to investigate the
predictions of $V_{\rm low-k}$ for the ground-state properties of light nuclei. For
that purpose we employ the calculation scheme, which has recently been
discussed in ref.~ \cite{khaga2} and apply it for the nucleus $^{16}$O. 
This scheme is an extension of the conventional BHF approximation for finite 
nuclei. It assumes a model space, which in our example is defined in terms of
configurations of nucleons occupying oscillator states up to the $1p0f$ shell.
Correlations outside this model space, which we would like to call short-range
correlations, are treated within the framework of the BHF approximation. 
The long-range correlations from configurations inside the model space are
treated within the framework of the Green's function approach, assuming a
self-energy which accounts for 2-particle - 1-hole and 2-hole - 1-particle
admixtures.  

This Green's function approach (for further details see ref.~ \cite{khaga2}) yields
the spectral function of the hole-state distribution $S_{\tau lj}^h(\omega)$,
which is the probability to remove a nucleon with quantum numbers $\tau$ for
isospin, $l$ for orbital angular momentum  and $j$ for the total angular
momentum by absorption of an energy $-\omega$. From this spectral function we
can easily determine the total occupation probability
\begin{equation}
\rho_{\tau lj} = \int_{-\infty}^{\varepsilon_F} d\omega \, S_{\tau
lj}^h(\omega)\,\label{eq:ocuu}
\end{equation}
the mean single-particle energy of the distribution function
\begin{equation}
\varepsilon_{\tau lj} = \int_{-\infty}^{\varepsilon_F} d\omega \, \omega\,
S_{\tau lj}^h(\omega)\,\label{eq:epsi}
\end{equation}
and the total energy 
\begin{equation}
E = \sum_{\tau l j} \frac{(2j+1)}{2} \int_{-\infty}^{\varepsilon_F} d\omega \,
S_{\tau lj}^h(\omega_{\tau  lj})\left\langle \frac{p^2}{2m} +
\omega_{\tau lj}\right\rangle \,.\label{eq:koltun}
\end{equation}

Results for these observables are listed in table~\ref{tab1} using the bare 
CD-Bonn interaction along with
 $V_{\rm low-k}$ derived from the CD-Bonn potential with
$\Lambda$ = 2 fm$^{-1}$ and 3 fm$^{-1}$. For the interactions $V_{\rm low-k}$ we
have performed calculations in which the effects of short-range correlations are
ignored (labeled HF) and taken into account (labeled BHF).

The BHF calculations employing $V_{\rm low-k}$ yield results which are very close 
to the corresponding ones for the bare interaction. The agreement is better for
the larger value of the cut-off $\Lambda$. The BHF calculations using 
$V_{\rm low-k}$ yield energies which are more attractive than the corresponding
result for the bare interaction. This can be explained in the same way as for
the case of nuclear matter: $V_{\rm low-k}$ accounts for correlations originating
from states above the cut-off as they occur for two free nucleons. The
dispersive quenching effect in the $G$-matrix calculated for the bare
interaction, however, reduces these attractive correlation contributions. We
find that this effect is rather small for finite nuclei, comparable to the same
effect in nuclear matter at small densities (see Fig.~\ref{fig1}).

The predictions from calculations using $V_{\rm low-k}$ in the HF approach
yield energies which are less attractive than those from the bare interaction.
This is due to the fact that they do not account for short-range correlations
originating from configurations outside the model space (up to $1p0f$
shell) but only below the cut-off. Nevertheless, 
the results obtained for $\Lambda$ = 2 fm$^{-1}$ are quite reasonable. 

The predictions for the occupation probabilities are larger for the calculations
using $V_{\rm low-k}$ than for the bare interaction $V$. This is due to the fact
that the HF calculations using $V_{\rm low-k}$ only account for a depletion of the
occupation due to long-range correlations within the $1p0f$ model space, while
the corresponding BHF calculations include effects of correlations only up to
the cut-off $\Lambda$. One may observe, however, that the BHF calculation using
$V_{\rm low-k}$ with $\Lambda$ = 3 fm$^{-1}$ yields results for the occupation which
are only slightly above those derived from the bare interaction. 

\begin{figure}[ht]
\begin{center}
\epsfig{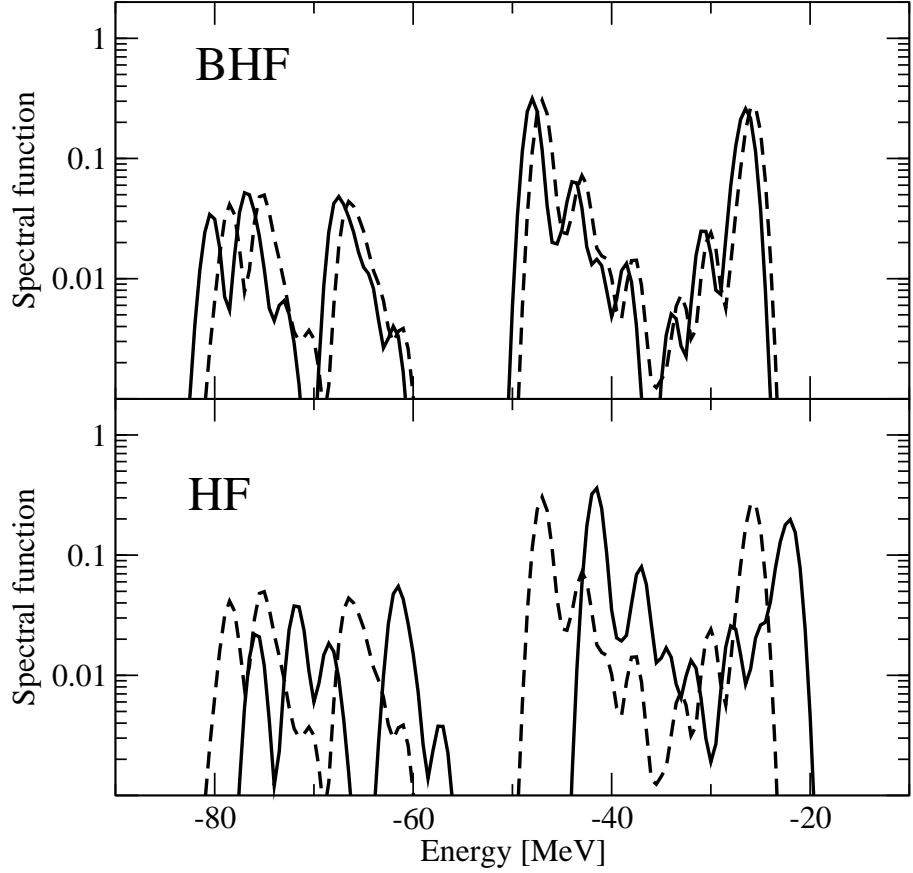}
\end{center}
\caption{The spectral function for proton hole strength in $^{16}$O in the
$s_{1/2}$ channel for the BHF (upper panel) and the HF (lower panel) 
approximations. The spectral distribution shown by the solid line
has been calculated for $V_{\rm low-k}$ with $\Lambda$ = 2 fm$^{-1}$.
For comparison, we show the results derived from the BHF approximation using 
the bare interaction (dashed lines)\protect \cite{khaga2}. 
\label{fig3}}
\end{figure}

The reliability of $V_{\rm low-k}$ in low-energy nuclear structure calculations of
finite nuclei is also supported by the results for the spectral function
displayed in Fig.~\ref{fig3}. The structure of the spectral function derived
from the bare interaction, represented by the dashed line in both parts of the
figure, is nicely reproduced by the spectral function derived from $V_{\rm low-k}$.
This is true in particular for the BHF approach, displayed by the solid line in
the upper panel, but also holds for the HF approximation, which is shown
in the lower panel. The main deficiency is a shift of the strength to larger
energies $\omega$, a feature which we have already outlined above in discussing
the mean single-particle energies shown in table~\ref{tab1}.   

\section{Pairing and Residual Interaction in Nuclear Matter}

\begin{figure}[ht]
\begin{center}
\epsfig{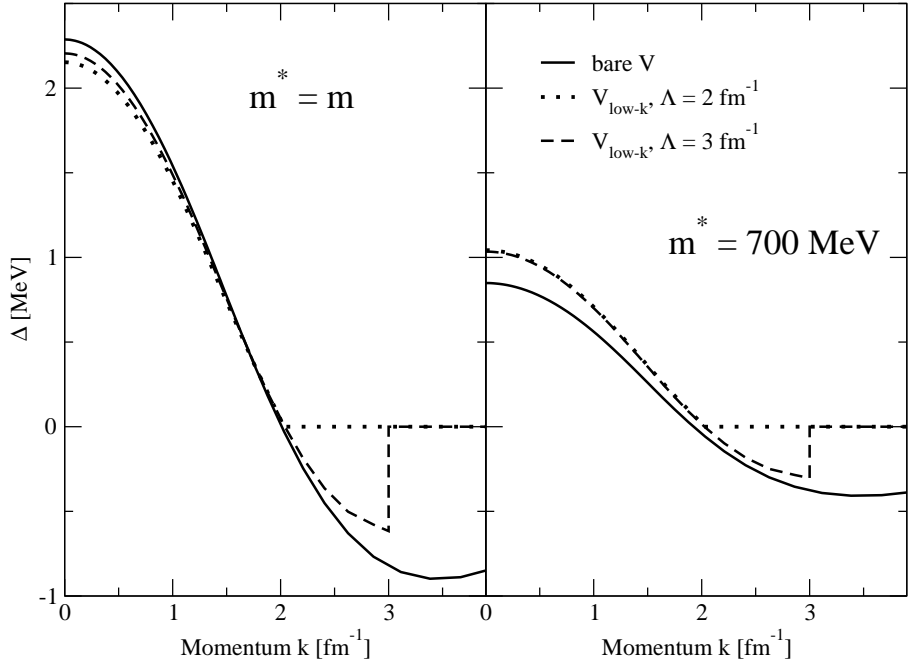}
\end{center}
\caption{The pairing gap $\Delta (k)$ for 
neutron-neutron pairing in the $^1S_0$
channel for neutron matter with a Fermi momentum $k_F$ = 1.36 fm$^{-1}$ as
function of the nucleon momentum $k$. The results obtained for the bare CD-Bonn
interaction (solid lines) are compared to those derived from
$V_{\rm low-k}$ using cut-offs $\Lambda= 2$ fm$^{-1}$ 
and 3 fm$^{-1}$. The left and right panels show the results 
for non-renormalized single-particle spectrum ($m^*=m$) and 
for the renormalized spectrum with an  effective mass $m^*$ = 700 MeV,
respectively. 
\label{fig4}}
\end{figure}

Up to this point we have focused the discussion on 
observables, which are characteristic for the ground state of nuclear systems. In
this section we would like to address the question how reliable is the use of
$V_{\rm low-k}$ for observables related to the low-energy excitation spectrum. As
the first example we consider the evaluation of the pairing gap in nuclear
matter. In particular we take the example of neutron-neutron pairing in the 
$^1S_0$ channel for neutron matter at $k_F$ = 1.36 fm$^{-1}$. This
pairing gap is determined from the solution of the gap equation (see e.g.
 \cite{baldo,elgar})

\begin{equation}
\Delta(k)= -\frac{2}{\pi} \int_0^\infty dk'\,k'^2\, V(k,k') \frac{\Delta(k')}
{2\sqrt{\left(\varepsilon_k' - \varepsilon_F\right)^2 + \Delta(k')^2}}\,.
\label{eq:gap}
\end{equation}
Here we will again parametrize the single-particle spectrum by
eq.~(\ref{eq:mstar}) and consider the cases $m^*=m$ and $m^*$ = 700 MeV.
For a realistic interaction, one has to integrate up to rather large values of
$k'$ before a convergent result is obtained. Corresponding results for
$\Delta(k)$ for the bare CD-Bonn interaction are displayed in
Fig.~\ref{fig4}. For the $V_{\rm low-k}$ interactions, 
the integral in (\ref{eq:gap}) can
be constrained to values of $k'$ below the cut-off $\Lambda$.

The agreement for the resulting gap functions is very good for momenta below
the cut-off $\Lambda$ and a single-particle spectrum with an effective 
mass identical to the bare mass. In this case the use of  $V_{\rm
low-k}$  in the gap equation corresponds to the use of the effective 
model-space interaction of ref.  \cite{elgar}. 
However, if we use a single-particle spectrum  renormalized by the
medium, the results obtained from $V_{\rm low-k}$ differ in a more significant way
from the corresponding result obtained for the bare CD-Bonn interaction. If,
e.g. we use an effective mass $m^*$ of 700 MeV (see right panel of
Fig.~\ref{fig4}), $V_{\rm low-k}$ tends to overestimate the value for the gap
$\Delta$ as it does not account for the reduction of the contributions from high
lying NN states, which is due to the decrease of the effective mass. This
feature is identical to the effect of the dispersive quenching of correlations
discussed above.

Finally, we would like to consider the residual interaction for low-energy
excitations as characterized by the Landau parameters. It
is not our goal to employ a very sophisticated calculational scheme for the
Landau parameters. This would involve in particular a careful analysis of the
contributions due to the induced interaction terms \cite{babu,wim2,schwen}. 
Our aim here is
to compare the results from the bare interaction $V_{NN}$, the corresponding
$G$-matrix and those derived from $V_{\rm low-k}$, in order to see to which extent
$V_{\rm low-k}$ accounts for the short-range correlation effects in a nuclear
system. For that purpose, we determine the Landau parameters either from the bare
interaction or the $G$-matrix \cite{wim1}.

The results of our calculations of 
the Landau parameters are listed in table~\ref{tab2}. Beside those
parameters we also list the effective mass determined by
\begin{equation}
\frac{m^*}{m} = 1 + \frac{1}{3} f_1 \label{eq:mland}
\end{equation}
and the coefficient $\beta$ for the symmetry energy
\begin{equation}
\beta = \frac{1}{3}\frac{k_F^2}{2m^*} \left( 1 + f'_0\right)\,.\label{eq:sym}
\end{equation}
Comparing the various approximations with our ``best'' result derived from the
$G$-matrix of the CD-Bonn potential, one finds that the bare interaction
$V_{\rm low-k}$ with a small cut-off parameter yields a good approximation if one
would like to restrict oneself 
to the bare interaction. If an iteration is carried out 
in the particle-particle channels for momenta up to the cut-off
$\Lambda$, one obtains the best results using a larger cut-off.

\section{Conclusions}
The aim of this study was to explore to which extent the effects of
short-range correlations, which are due to excited states above a cut-off
parameter $\Lambda$, are accounted for by the effective interaction
$V_{\rm low-k}$, which has recently been constructed
 \cite{bogner01,bogner02,bogschwenk}. This effective interaction accounts for
short-range correlations of two-nucleons interacting in the vacuum;  the question is
whether this is a good approximation also for the corresponding correlations of
nucleons interacting in a nuclear medium. To study this question we considered
various nuclear structure observables.

Inspecting the equation of state calculated for nuclear matter one may
conclude that $V_{\rm low-k}$ yields a good approximation for BHF calculations at
small densities. In this regime even a Hartree-Fock calculation leads to
reasonable results, if $\Lambda$ is small enough. At higher densities, however,
$V_{\rm low-k}$ produces too much binding as it does not account for the dispersive
quenching effects, which are taken into account e.g.~in the 
$G$-matrix. Therefore calculations using $V_{\rm low-k}$ cannot reproduce the
saturation feature of nuclear matter.

The $V_{\rm low-k}$ interaction is well suited for studies of finite
nuclei like $^{16}$O, which is again a system of small density. In this case we
obtain reasonable agreement for the calculated energies, but also for the
spectral function for nucleon removal. 

Furthermore, we demonstrated that the use of $V_{\rm low-k}$ is also very
efficient in calculating quantities like the pairing gap in infinite nuclear
matter. The agreement with the exact result is very good if one may 
replace the single-particle energy above the cut-off by pure kinetic energy.
Deviations are observed if this is not the case. These deviations are again 
related to the dispersive quenching mechanism mentioned above.

In conclusion we would like to emphasize that $V_{\rm low-k}$ can be a very useful
tool for low-energy nuclear structure calculations. However, one should be
cautious if the observable of interest
is sensitive to the single-particle spectrum at
energies above the cut-off. This is a problem in particular for high densities.

\begin{ack} We thank Prof.~T.T.S.~Kuo for supplying the matrix
elements of $V_{\rm low-k}$ and for many stimulating discussions.
This work was supported in part by  the DFG (SFB 382) and through 
a stipend of the DAAD.
\end{ack}


\begin{table}[ht]
\begin{center}
\begin{tabular}{c|rr|rrr|r}
& \multicolumn{2}{c}{HF} & \multicolumn{3}{c}{BHF} & \\
& $\Lambda$=2 fm$^{-1}$ & $\Lambda$=3 fm$^{-1}$ & $\Lambda$=2 fm$^{-1}$ 
& $\Lambda$=3 fm$^{-1}$&  bare V & Exp\\
\hline
$\varepsilon_{s1/2}$ & -39.28 & -31.31 & -45.23 & -44.74 & -44.00 
& -40 $\pm$ 8 \\
$\varepsilon_{p_3/2}$ & -18.32 & -12.34 & -25.07 & -24.83 & -24.29 
& -18.45 \\
$\varepsilon_{p_1/2}$ & -15.01 & -9.01 & -21.97 & -21.81 & -21.26 
& -12.13 \\
\hline
$\rho_{s1/2}$ & 0.964 & 0.964 & 0.906 & 0.880 & 0.867  & \\
$\rho_{p3/2}$ & 0.928 & 0.936 & 0.844 & 0.828 & 0.823  & \\
$\rho_{p1/2}$ & 0.912 & 0.912 & 0.832 & 0.815 & 0.802  & \\
\hline
E/A & -4.96 & -1.69 & -8.22 & -8.08 & -7.78 & -7.98 \\
\end{tabular}
\caption{\label{tab1} Single-particle energies ($\varepsilon_i$) and occupation
probabilities ($\rho_i$) for protons in $^{16}$O and the total energy per
nucleon (E/A) as calculated from various approximations and interactions are
compared to the experimental data. The CD Bonn interaction has been used (bare
V) and the effective interactions $V_{\rm low-k}$ derived from it assuming a cut-off
$\Lambda$ of 2.0 fm$^{-1}$ and 3.0 fm$^{-1}$. The effects of short-range
correlations have been ignored (HF) or taken into account by means of the BHF
approximation. All energies are given in
MeV.}
\end{center}
\end{table}

\begin{table}[ht]
\begin{center}
\begin{tabular}{c|rrr|rrr}
& \multicolumn{3}{c}{$V$} & \multicolumn{3}{c}{$G$} \\
& $\Lambda$=2 fm$^{-1}$ & $\Lambda$=3 fm$^{-1}$ & bare V &$\Lambda$=2 fm$^{-1}$ 
& $\Lambda$=3 fm$^{-1}$&  bare V \\
\hline
$f_0$ & -1.819 & -1.315 & -0.465 & -2.003 & -1.807 & -1.601 \\
$f'_0$ & 0.502 & 0.235 & 0.072 & 0.614 & 0.492 & 0.422 \\
$g_0$ & 0.250 & 0.362 & 0.226 & 0.211 & 0.279 & 0.276 \\
$g'_0$ & 0.970 & 0.899 & 0.709 & 0.998 & 0.973 & 0.929 \\
$h_0$ & 1.788 & 1.759 & 1.618 & 1.753 & 1.717 & 1.694 \\
$h'_0$ & -0.596 & -0.631 & -.581 & -.588 & -.620 & -.613 \\
\hline
$m^*/m$ & 0.702 & 0.684 & 0.620 & 0.696 & 0.688 & 0.676 \\
$\beta$ [MeV] & 28.98 & 24.46 & 23.43 & 31.40 & 29.35 &  28.48 \\
\end{tabular}
\caption{\label{tab2} Landau parameters, effective mass and the coefficient for
the symmetry energy $\beta$ as calculated from the interaction $V$ and the
corresponding $G$ matrix in nuclear matter at a Fermi momentum $k_F$ of 1.36
fm$^{-1}$.  The CD Bonn interaction has been used (bare
V) and the effective interactions $V_{\rm low-k}$ derived from it assuming a cut-off
$\Lambda$ of 2.0 fm$^{-1}$ and 3.0 fm$^{-1}$. The Landau parameters are
normalized with respect to the density of states at the Fermi momentum using
an effective mass as listed.}
\end{center}
\end{table}

\end{document}